\documentclass[11pt]{article}
\usepackage{amssymb,amsmath,graphicx,cite}
\begin{document}
\begin{titlepage}
\vspace{.5in}
\begin{flushright}
November 2009\\  
\end{flushright}
\vspace{.5in}
\begin{center}
{\Large\bf
 The Small Scale Structure of Spacetime}\\   
\vspace{.4in}
{S.~C{\sc arlip}\footnote{\it email: carlip@physics.ucdavis.edu}\\
       {\small\it Department of Physics}\\
       {\small\it University of California}\\
       {\small\it Davis, CA 95616}\\{\small\it USA}}
\end{center}

\vspace{.5in}
 
\begin{abstract}
Several lines of evidence hint that quantum gravity at very small 
distances may be effectively two-dimensional.  I summarize the
evidence for such ``spontaneous dimensional reduction,'' and 
suggest an additional argument coming from the strong-coupling 
limit of the Wheeler-DeWitt equation.  If this description proves 
to be correct, it suggests a fascinating relationship between 
small-scale quantum spacetime and the behavior of cosmologies 
near an asymptotically silent singularity.
\end{abstract}
\end{titlepage}
\addtocounter{footnote}{-1}

Stephen Hawking and George Ellis prefaced their seminal book, 
\emph{The Large Scale Structure of Space-Time}, with the
explanation that their aim was to understand spacetime ``on
length-scales from $10^{-13}$ cm, the radius of an elementary 
particle, up to $10^{28}$ cm, the radius of the universe''
\cite{HawkingEllis}.  While many deep questions remain,
ranging from cosmic censorship to the actual topology of our
universe, we now understand the basic structure of spacetime at 
these scales: to the best of our ability to measure such a thing, 
it behaves as a smooth (3+1)-dimensional Riemannian manifold.  

At much smaller scales, on the other hand, the proper description 
is far less obvious.  While clever experimentalists have managed 
to probe some features down to distances close to the Planck 
scale \cite{Mattingly}, for the most part we have neither direct 
observations nor a generally accepted theoretical framework
for describing the  very small-scale structure of spacetime.  Indeed, 
it is not completely clear that ``space'' and ``time'' are even the 
appropriate categories for such a description.

But while a complete quantum theory of gravity remains elusive,  
we do have fragments: approximations, simple models, and pieces
of what may eventually prove to be the correct theory.  None of 
these fragments is reliable by itself, but when they agree with 
each other about some fundamental property of spacetime, we 
should consider the possibility that they are showing us something 
real.  The thermodynamic properties of black holes, for example,
appear so consistently that it is reasonable to suppose that they 
reflect an underlying statistical mechanics of quantum states.

Over the past several years, evidence for another basic feature of
small-scale spacetime has been accumulating: it is becoming 
increasingly plausible that spacetime near the Planck scale is 
effectively two-dimensional.  No single piece of evidence for this 
behavior is in itself very convincing, and most of the results are
fairly new and tentative.  But we now have hints from a number of 
independent calculations, based on different approaches to quantum 
gravity, that all point in the same direction.  Here, I will summarize 
these clues, provide a further piece of evidence in the form of a 
strong-coupling approximation to the Wheeler-DeWitt equation, 
and discuss some possible implications.

\section{Spontaneous dimensional reduction?}

Hints of short-distance ``spontaneous dimensional reduction'' in
quantum gravity come from a number of places.  Here I will
review some of the highlights:\\

\noindent{\bf Causal Dynamical Triangulations}\\[-1.5ex]

As we have learned from quantum chromodynamics---and from our
colleagues in condensed matter physics---lattice approximations to
the Feynman path integral can give us valuable information about
the nonperturbative behavior of theories that may otherwise be
extremely difficult to analyze.  Lattice approximations to quantum
gravity are not quite typical: in contrast to QCD, where fields live on 
a fixed lattice, gravity \emph{is} the lattice \cite{Regge}, which 
forms a discrete approximation of a continuous spacetime geometry.  
Despite this difference, though, we might hope that a suitable 
lattice formulation could tell us something important about 
quantized spacetime.

\begin{figure}
\includegraphics[width=1.96in]{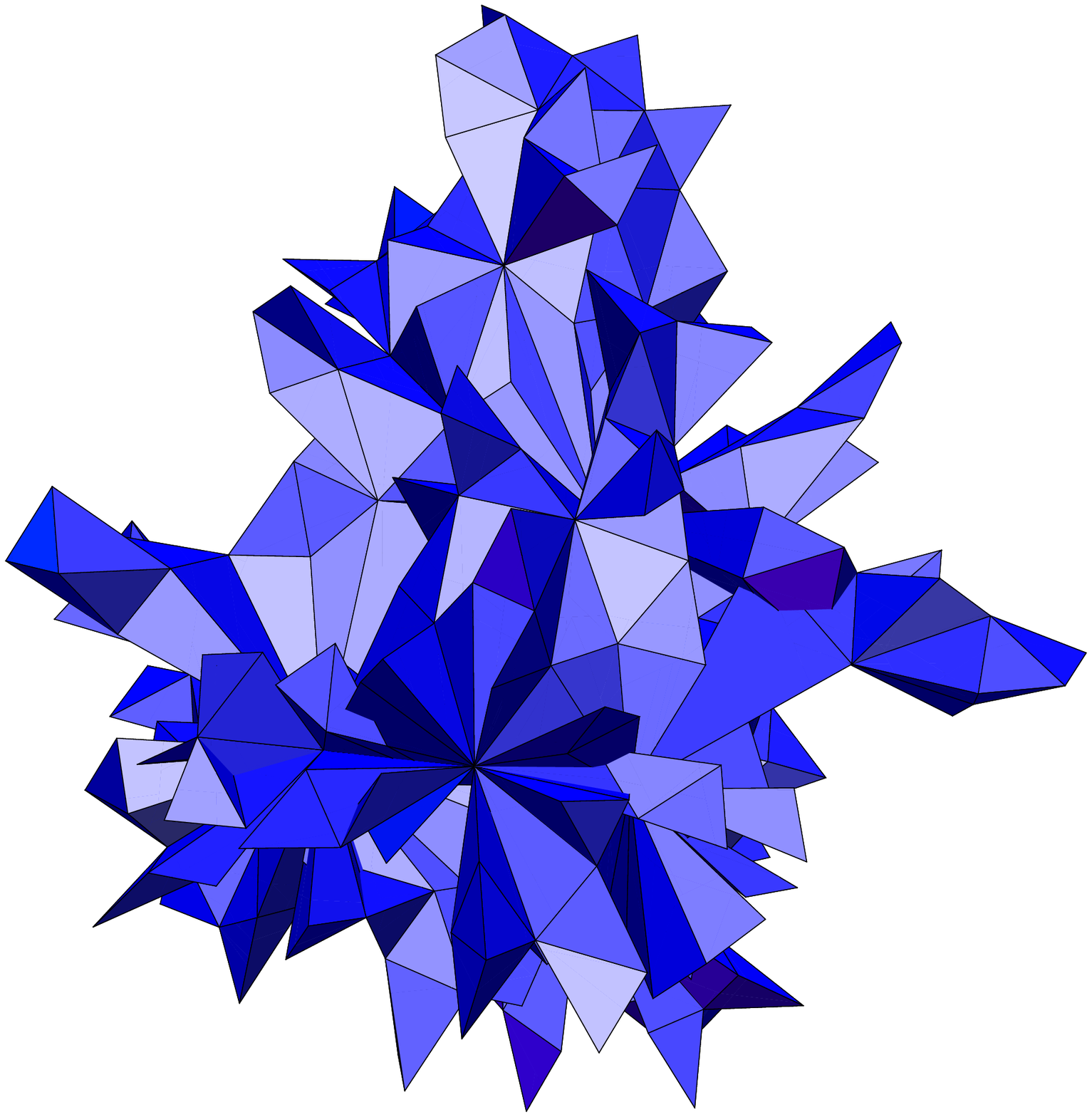}
\includegraphics[width=4in]{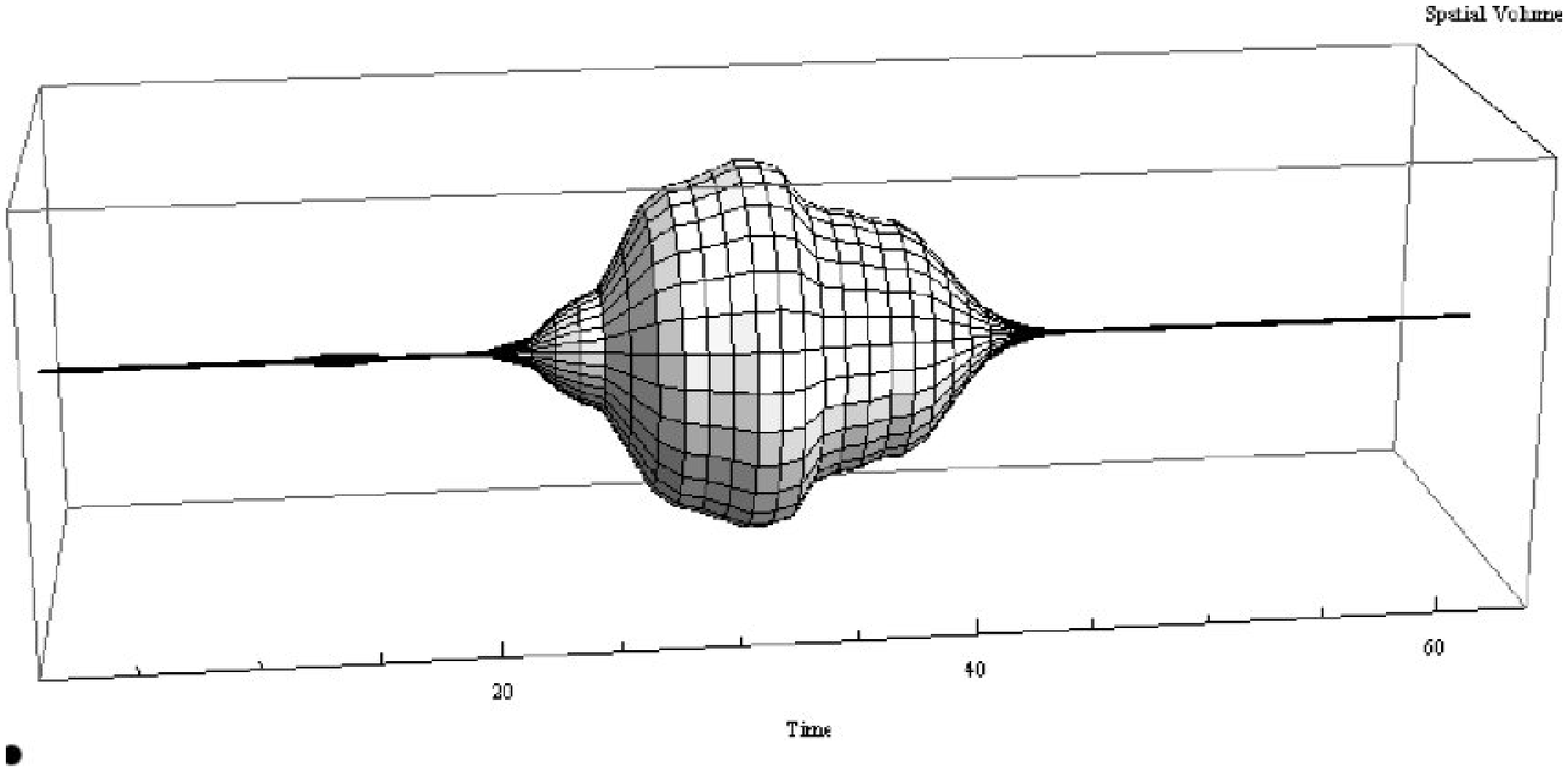} 
\caption{A spatial slice and a typical history contributing to the 
causal dynamical triangulations path integral (R.\ Kommu and M.\ Sachs, 
UC Davis)}
\label{CDT}
\end{figure}

The idea of combining Regge calculus with Monte Carlo methods to
evaluate the gravitational path integral on a computer dates back to  
1981 \cite{Williams}.  Until fairly recently, though, no good continuum
limit could be found.  Instead, the simulations typically yielded two
unphysical phases, a ``crumpled'' phase with very high Hausdorff 
dimension and a two-dimensional ``branched polymer'' phase \cite{Loll}.  
The causal dynamical triangulation program of Ambj{\o}rn, 
Jurkiewicz, and Loll \cite{AJL,AJL2,AJL3} adds a crucial new ingredient, 
a fixed causal structure in the form of a prescribed time-slicing.  By 
controlling fluctuations in topology, this added structure suppresses 
the undesirable phases, and appears to lead to a good four-dimensional 
continuum picture.   Results so far are very promising; in particular, 
the cosmological scale factor appears to have the correct semiclassical 
behavior \cite{AJL3,AJL4}.  Figure \ref{CDT} illustrates a typical time 
slice and a typical history contributing to the path integral in a simulation 
developed at UC Davis \cite{Kommu}.    

A crucial question for any such microscopic approximation is whether it 
can genuinely reproduce the four-dimensional structure we observe at 
``normal'' distances.  This is a subtle issue, which cannot be answered by
merely looking at particular contributions to the path integral.  As
a first step, we need a definition of ``dimension'' for a discrete structure 
that may be very non-manifold-like at short distances.   One natural 
choice---although by no means the only one---is the spectral dimension 
\cite{spec}, the dimension as seen by a diffusion process or a random 
walker.  

The basic idea of the spectral dimension is simple.  In any structure on
which a random walk can be defined, the associated diffusion process
will gradually explore larger and larger regions of the structure.  The
more dimensions available for the random walk to explore, though, 
the longer this diffusion will take.  Quantitatively, diffusion from an 
initial position $x$ to a final position $x'$ on a manifold $M$ may be 
described by a heat kernel $K(x,x',s)$ satisfying
\begin{equation}
\left(\frac{\partial\ }{\partial s} - \Delta_x\right)K(x,x';s) =0,
\qquad \hbox{with \quad $K(x,x',0) = \delta(x-x')$} ,
\label{Ca1}
\end{equation}
where $\Delta_x$ is the Laplacian on $M$ at $x$, and $s$ is a measure 
of the diffusion time.  Let $\sigma(x,x')$ be Synge's world function
\cite{Synge}, one-half the square of the geodesic distance between $x$ 
and $x'$.  Then on a manifold of dimension $d_S$, the heat kernel 
generically behaves as
\begin{equation}
K(x,x';s) \sim (4\pi s)^{-d_S/2} e^{-\sigma(x,x')/2s}
    \left( 1 + \mathcal{O}(s)\right)
\label{Ca2}
\end{equation}
for small $s$.  In particular, the return probability $K(x,x,s)$ is
\begin{equation}
K(x,x;s) \sim (4\pi s)^{-d_S/2} .
\label{Ca3}
\end{equation}

For any structure on which a diffusion process can be defined, be it a 
manifold or not, we can now use equation (\ref{Ca3}) to define an 
effective dimension $d_S$, the spectral dimension.  On a lattice, in 
particular, we can determine the spectral dimension by directly 
simulating random walks.  In the causal dynamical triangulation 
program, the ensemble average over the histories contributing to 
the path integral then gives a quantum spectral dimension.  The 
results of such simulations yield a spectral dimension of $d_S=4$ at 
large distances \cite{AJL3,spec}.  This is a promising sign, indicating 
the recovery of four-dimensional behavior.  Similarly, (2+1)-dimensional
causal dynamical triangulations yield a large-distance spectral
dimension of $d_S=3$ \cite{Kommu,Henson}.

At short distances, though, the result is dramatically different. In
both 3+1 dimensions and 2+1 dimensions, the small-scale spectral 
dimension falls to $d_S=2$.  This is the first, and perhaps the clearest, 
sign of spontaneous dimensional reduction at short distances.

As noted above, $d_S$ is by no means the only definition of a generalized 
dimension, and one may worry about reading too much significance into 
this result.  Note, though, that the heat kernel has a special significance 
in quantum field theory: propagators of quantum fields may be obtained 
as Laplace transforms of appropriate heat kernels.  For a scalar field,
in particular, the propagator is determined by the heat kernel (\ref{Ca1}),
and the behavior of the spectral dimension implies a structure
\begin{equation}
G(x,x') \sim \int_0^\infty \!ds\, K(x,x';s) \sim \left\{\begin{array}{cl}
   \sigma^{-2} \quad & \hbox{at large distances}\\
   \ln\sigma \quad & \hbox{at small distances.} \end{array}\right. 
\label{Ca4}
\end{equation}
The logarithmic short-distance behavior is the standard result for a
two-dimensional conformal field theory.  If one probes short distances 
with a quantum field, the field will thus act as if it lives in an effective 
dimension of two.\\

\noindent{\bf Renormalization Group Analysis}\\[-1.5ex]

General relativity is nonrenormalizable: conventional perturbative
quantum field theory techniques yield an infinite number of higher
derivative counterterms, each with its own coupling constant.
Nevertheless, the renormalization group \emph{flow} of these
coupling constants may provide us with valuable information about 
quantum gravity.   In particular, a renormalization group analysis, 
even if incomplete or truncated, may offer a method for probing 
the theory at high energies and short distances.

One particularly dramatic possibility, first suggested by Weinberg
\cite{Weinberg}, is that general relativity may be ``asymptotically 
safe.''  Consider the full effective action for conventional gravity, 
with its infinitely many coupling constants.  The renormalization 
group describes the dependence of these constants on energy scale, 
and in principle allows us to compute the high-energy/small-distance 
(``ultraviolet'') couplings in terms of their low-energy/large-distance
(``infrared'') values.  Under the renormalization group flow toward
high energies, some of these constants may blow up, indicating that 
the description has broken down and that new physics is needed.  An 
alternative possibility, though, is that the coupling constants might 
remain finite and flow to an ultraviolet fixed point.  In that case, the 
theory would continue to make sense down to arbitrarily short 
distances.  If, in addition, the critical surface---the space of such UV 
fixed points---were finite dimensional, the long-distance coupling 
constants would be determined by a finite number of short-distance
parameters: not quite renormalizability, but perhaps almost as good.

Distinguishing among these possibilities (and others) is extremely
difficult, and we are still a long way from knowing whether quantum 
general relativity is asymptotically safe.  But there is growing 
evidence for a UV fixed point, coming from various truncations of 
the effective action and from exact calculations in dimensionally 
reduced models \cite{Reuter,Litim,Nieder}.  For our present purposes, 
the key result of these calculations is that field operators acquire 
large anomalous dimensions; that is, under a change in mass scale, 
they scale differently than one would expect from dimensional analysis 
based on their classical ``engineering dimension.''  In fact, these
operators scale precisely as one would expect for the corresponding  
quantities in a \emph{two-dimensional} field theory \cite{Reuter}.  
Moreover, the spectral dimension near the putative fixed point can
be computed using field theoretical techniques, and the result is
again $d_S=2$ \cite{Reuter2}.

There is, in fact, a fairly general argument that if quantum gravity
is asymptotically safe, it must be effectively two-dimensional at very
short distances \cite{Nieder,Litimb}.  Consider the dimensionless
coupling constant $g_N(\mu) = G_N\mu^{d-2}$, where $G_N$ is
Newton's constant and $\mu$ is the mass scale that appears in the
renormalization group flow.  Under this flow,
\begin{equation}
\mu\frac{\partial g_N}{\partial\mu} 
   = [d-2+\eta_N(g_N,\dots)] g_N  ,
\label{Ca5}
\end{equation}
where the anomalous dimension $\eta_N$ depends upon both $g_N$ and 
any other dimensionless coupling constants in the theory.  Evidently
a free field (or ``Gaussian'') fixed point can occur at $g_N=0$.  For 
an additional non-Gaussian fixed point $g_N^*$  to be present, though,
the right-hand side of (\ref{Ca5}) must vanish: $\eta_N(g_N^*,\dots) = 2-d$.

But the momentum space propagator for a field with an anomalous 
dimension $\eta_N$ has a momentum dependence $(p^2)^{-1 + \eta_N/2}$.  
For $\eta_N = 2-d$, this becomes $p^{-d}$, and the associated position 
space propagator depends logarithmically on distance.  As I noted earlier,  
such a  logarithmic dependence is characteristic of a two-dimensional 
conformal field.  A variation of this argument shows that arbitrary matter 
fields interacting with gravity at a non-Gaussian fixed point exhibit a
 similar two-dimensional behavior \cite{Nieder}.\\

\noindent{\bf Loop quantum gravity}\\[-1.5ex]

A third hint of short-distance dimensional reduction comes from the area 
spectrum of loop quantum gravity \cite{Modesto}.  States in this proposed
quantum theory of gravity are given by spin networks, graphs with edges 
labeled by half-integers $j$ (which represent holonomies of a connection) 
and vertices labeled by $\mathrm{SU}(2)$ intertwiners, that is, generalized
Clebsch-Gordan coefficients.  Given such a state, the area operator for a 
surface counts the number of spin network edges that puncture the surface, 
with each such edge contributing an amount
$$A_j \sim \ell_p^2\sqrt{j(j+1)},$$
where 
\begin{equation}
\ell_p = \sqrt{\frac{\hbar G}{c^3}}
\label{Ca5a}
\end{equation}
is the Planck length.  

While area and volume operators in loop quantum gravity are well
understood, it has proven rather difficult to define a length operator.
But since $j$ is, crudely speaking, a quantum of area, one can (equally 
crudely) think of $\sqrt{j}$ as a sort of quantum of length.  If we 
therefore define a length $\ell_j = \sqrt{j}\ell_p$, we can rewrite the 
spectrum as
\begin{equation}
A_j \sim \sqrt{\ell_j^2(\ell_j^2 + \ell_p^2)} \sim \left\{\begin{array}{cl}
   \ell_j^2 \quad & \hbox{for large areas}\\
   \ell_p\ell_j \quad & \hbox{for small areas.} \end{array}\right. 
\label{Ca6}
\end{equation}
Like the propagator (\ref{Ca4}), this spectrum undergoes a change 
in scaling at small distances.  Indeed, one can define a scale-dependent
effective metric that reproduces the behavior (\ref{Ca6}), and use it to
compute an effective spectral dimension.  One again finds a dimension 
that decreases from four at large scales to two at small scales.\\

\noindent{\bf High temperature strings}\\[-1.5ex]

A fourth piece of evidence for small-scale dimensional reduction comes 
from the behavior of string theory at high temperatures.  At a critical 
temperature, the Hagedorn temperature, the string theory partition 
function diverges, and the theory (probably) undergoes a phase transition.  
As early as 1988, Atick and Witten discovered that at temperatures far 
above the Hagedorn temperature, string theory has an unexpected 
thermodynamic behavior \cite{Atick}: the free energy in a volume $V$ 
varies with temperature as
\begin{equation}
F/VT \sim T .
\label{Ca7}
\end{equation}
For a field theory in $d$ dimensions, in contrast, $F/VT\sim T^{d-1}$.
So even though string theory lives in 10 or 26 dimensions, at high 
temperatures it behaves thermodynamically as if spacetime were 
two-dimensional.\\

\noindent{\bf Anisotropic scaling models}\\[-1.5ex]

A fifth sign of short-distance spontaneous dimensional reduction in 
quantum gravity comes from ``Ho{\v r}ava-Lifshitz'' models \cite{Horava}.  
These are  new models of gravity that exhibit anisotropic scaling, that 
is, invariance under constant rescalings 
$${\bf x}\rightarrow b{\bf x},\ t\rightarrow b^3t .$$
This scaling property clearly violates Lorentz invariance, breaking 
the symmetry between space and time and picking out a preferred time
coordinate.  In fact, it is this breaking of Lorentz invariance that makes 
the models renormalizable: the field equations may now contain many 
spatial derivatives, leading to high inverse powers of spatial momentum 
in propagators that can tame loop integrals, while keeping only second 
time derivatives, thus avoiding negative energy states or negative norm 
ghosts.  This might seem an unlikely way to quantize gravity---and 
indeed, these models may face serious low-energy problems \cite{Blas}%
---but the hope is that Lorentz invariance might be recovered and 
conventional general relativity restored at low energies and large 
distances.

Ho{\v r}ava has calculated the spectral dimension in such models
\cite{Horava2}, and again finds $d_S=2$ at high energies.  In this
case, the two-dimensional behavior can be traced to the fact that
the propagators contain higher inverse powers of momentum; the 
logarithmic dependence on distance comes from integrals of the form 
$$\int \frac{d^4p}{p^4}e^{ip\cdot(x-x')} .$$ 
Whether this is ``real'' dimensional reduction becomes a subtle 
matter, which may depend on how one operationally defines dimension.  
As before, though, the logarithmic behavior of propagators implies
that quantum fields used to probe the structure of spacetime will
act as if they are in a two-dimensional space.\\

\noindent{\bf Other hints}\\[-1.5ex]

Hints of two-dimensional behavior come from several other places 
as well.  In the causal set approach to quantum gravity, spacetime is 
taken to be fundamentally discrete, with a ``geometry'' determined
by the causal relationships among points.  In this setting, a natural 
definition of dimension is the Myrheim-Meyer dimension, which 
compares the number of causal relations between pairs of points 
to the corresponding number for points randomly sprinkled in a
$d_{M}$-dimensional Minkowski space (see, for instance, \cite{Rideout}).  
For a small enough region of spacetime, one might guess that the 
causal structure is generic, coming from a random causal ordering.
In that case, the Myrheim-Meyer dimension is approximately 2.38%
---not quite $2$, but surprisingly close \cite{Sorkin}.  

Dimensional reduction has also appeared in an analysis of quantum
field theory in a background ``foam'' of virtual black holes \cite{Crane},
although the effective dimension depends on the (unknown) black
hole distribution function.  A dimension of $2$ also appears in Connes' 
noncommutative geometrical description of general relativity 
\cite{Connes}; it is not clear to me whether this is related to the
dimensional reduction considered here.

\section{Strong coupling and small-scale structure}

Let us suppose these hints are really telling us something fundamental
about the small-scale structure of quantum gravity.  We then face
a rather bewildering question: which two dimension?  How can
a four-dimensional theory with no background structure or preferred
direction pick out two ``special'' dimensions at short distances?
To try to answer this question, it is worth looking at one more approach 
to Planck scale physics: the strong-coupling approximation to the 
Wheeler-DeWitt equation.

In the conventional Dirac quantization of canonical general relativity,
the configuration space variables are given by the spatial metric 
$g_{ij}$ on a time slice $\Sigma$, while their canonically conjugate 
momenta are related to the extrinsic curvature of the slice.  The 
Hamiltonian constraint, which expresses invariance under diffeomorphisms 
that deform the slice $\Sigma$, then acts on states $\Psi[g]$ to give 
the Wheeler-DeWitt equation \cite{DeWitt}
\begin{equation}
\left\{ 16\pi\ell_p^2G_{ijkl}
    \frac{\delta\ }{\delta g_{ij}} \frac{\delta\ }{\delta g_{kl}}
    - \frac{1}{16\pi\ell_p^2}\sqrt{g}\,{}^{(3)}\!R\right\}\Psi[g] = 0 
\label{Ca8}
\end{equation}
where
\begin{equation}
G_{ijkl} = \frac{1}{2}g^{-1/2}\left( g_{ik}g_{jl} + g_{il}g_{jk} -
  g_{ij}g_{kl}\right)
\label{Ca8a}
\end{equation}
is the DeWitt metric on the space of metrics.  The Wheeler-DeWitt 
equation is not terribly well-defined---the operator ordering is 
ambiguous, the product of functional derivatives must be regularized, 
and the wave functions $\Psi[g]$ should really be functions of spatial 
diffeomorphism classes of metrics---and it is not at all clear how to 
find an appropriate inner product on the space of solutions \cite{Carlip}.  
Nevertheless, the equation is widely accepted as a heuristic guide to 
the structure of quantum gravity.

Note now that spatial derivatives of the three-metric appear only in 
the scalar curvature term ${}^{(3)}\!R$ in (\ref{Ca8}).  As early as 
1976, Isham \cite{Isham} observed that this structure implied an 
interesting strong-coupling limit $\ell_p\rightarrow\infty$.  As the 
Planck length becomes large---that is, as we probe scales near or 
below $\ell_p$ \cite{Pilati,Maeda}---the scalar curvature term 
becomes negligible.  Neighboring spatial points thus effectively 
decouple, and the equation becomes ultralocal.  

The absence of spatial derivatives greatly simplifies the Wheeler-DeWitt 
equation, which  becomes exactly solvable.  Its properties in this limit 
have been studied extensively \cite{HPT,Teitelboim,Pilati2,Rovelli,Husain},
and preliminary attempts have been made to restore the coupling 
between neighboring points by treating the scalar curvature term as 
a perturbation \cite{Francisco,Salopek,Kirillov,Maeda}.  The same 
kind of perturbative treatment of the scalar curvature is important 
classically in a very different setting: it is central to the Belinskii-%
Khalatnikov-Lifshitz (BKL) approach to cosmology near a spacelike 
singularity \cite{BKL,HUR}.  We can therefore look to this classical 
setting for clues about the small-scale structure of quantum gravity.

To understand the physics of the strong-coupling approximation, it is 
helpful to note that the Planck length (\ref{Ca5a}) also depends on 
the speed of light.  In fact, this approximation can also be viewed as a 
small $c$ (``anti-Newtonian'') approximation.  As the Planck length 
becomes large, particle horizons shrink and light cones collapse to 
timelike lines, leading to the decoupling of neighboring points and 
the consequent ultralocal behavior \cite{Henneaux}.  In the completely 
decoupled limit, the classical solution at each point is a Kasner space,
\begin{align}
ds^2 =& dt^2 - t^{2p_1}dx^2 - t^{2p_2}dy^2 - t^{2p_3}dz^2 \\
\label{Ca9}
&\hbox{\small ($-\frac{1}{3}<p_1<0<p_2<p_3,\quad
p_1+p_2+p_3 = 1 = p_1^2+p_2^2+p_3^2$).}\nonumber
\end{align}
More precisely---see, for example, \cite{Helfer}---the general solution 
is an arbitrary $\mathrm{GL}(3)$ transformation of a Kasner metric.
This is still essentially a Kasner space, but now with arbitrary, not 
necessary orthogonal, axes.  

For large but finite $\ell_p$, the classical solution exhibits BKL 
behavior \cite{BKL,HUR}.  At any given point, the metric spends most 
of its time in a nearly Kasner form.  But as the metric evolves, the 
scalar curvature can grow abruptly.  The curvature term in the
Hamiltonian constraint---the classical counterpart of the Wheeler-DeWitt
equation---then acts as a potential wall,  causing a Mixmaster-like 
``bounce'' \cite{Misner} to a new Kasner solution with different axes 
and exponents.  In contrast to the ultralocal behavior at the strong-%
coupling limit, neighboring points are now no longer completely 
decoupled.  But the Mixmaster bounces are chaotic  \cite{Chernoff}, 
and the geometries at nearby points quickly become uncorrelated, 
with Kasner exponents occurring randomly with a known probability 
distribution \cite{Kirillov2}.

We can now return to the problem of dimensional reduction.  Consider
a timelike geodesic in Kasner space, starting at $t=t_0$ with a random 
initial velocity.  The geodesic equation is exactly integrable, and in the 
direction of decreasing $t$, the proper spatial distance traveled along 
each Kasner axis asymptotes to
\begin{equation}
\begin{array}{l} s_x \sim t^{p_1}\\
                                  s_y \sim 0\\
                                  s_z \sim 0 . \end{array}
\label{Ca10}
\end{equation}
Particle horizons thus shrink to lines, and geodesics effectively explore 
only one spatial dimension.  In the direction of increasing $t$, the 
results are similar, though a bit less dramatic:
\begin{equation}
\begin{array}{l} s_x  \sim t\\ 
                                  s_y  \sim t^{\max(p_2,1+p_1-p_2)}\\
                                  s_z  \sim t^{p_3} . \end{array}
\label{Ca11}
\end{equation}
Since $p_2$, $1+p_1-p_2$, and $p_3$ are all less than one, a random
geodesic again predominantly sees only one spatial dimension.

Since the heat kernel describes a random walk, one might expect this
behavior of Kasner geodesics to be reflected in the spectral dimension.  
The exact form of heat kernel for Kasner space is not known, but Futamase 
\cite{Futamase} and Berkin \cite{Berkin} have evaluated $K(x,x';s)$
in certain approximations.  Both find behavior of the form
\begin{equation}
K(x,x;s) \sim \frac{1}{4\pi s^2}\left[ 1 + \frac{a}{t^2}\,s + \dots \right] .
\label{Ca12}
\end{equation}
The interpretation of this expression involves an order-of-limits question.
For a fixed time $t$, one can always find $s$ small enough that the first
term in (\ref{Ca12}) dominates.  This is not surprising: the heat kernel is 
a classical object, and we are still looking at a setting in which the 
underlying classical spacetime is four-dimensional.  For a fixed return time 
$s$, on the other hand, one can always find a time $t$ small enough that 
the second term dominates, leading to an effective spectral dimension of 
two.  One might worry about the higher order terms in (\ref{Ca12}), 
which involve higher inverse powers of $t$ and might dominate at smaller
times.  But these terms do not contribute to the singular part of the
propagator (\ref{Ca4}); rather, they give terms that go as positive powers 
of the geodesic distance, and are irrelevant for short distance singularities, 
light cone behavior, and the like.  

One can investigate the same problem via the Seeley-DeWitt expansion
of the heat kernel \cite{DeWitt2,Gibbons,Vassilevich},
\begin{equation}
K(x,x,s)  \sim \frac{1}{4\pi s^2}\left( [a_0] + [a_1]s + [a_2]s^2 
   +\dots \right) .
\label{Ca13}
\end{equation}
The ``Hamidew coefficient'' $[a_1]$ is proportional to the scalar 
curvature, and vanishes for an exact vacuum solution of the field 
equations.  In the presence of matter, however, the scalar curvature
will typically increase as an inverse power of $t$ as $t\rightarrow0$
\cite{BKL}; this growth is slow enough to not disrupt the BKL behavior
of the classical solutions near $t=0$, but it will nevertheless give a
diverging contribution to $[a_1]$.

This short-distance BKL behavior of the strongly coupled Wheeler-DeWitt
equation may thus offer an explanation for the apparent dimensional 
reduction of quantum gravity at the Planck scale.   We now have a 
possible answer to the question, ``Which two dimensions?''  If this 
picture is right, the dynamics picks out an essentially random timelike 
plane at each point.  This choice, in turn, is reflected in the behavior 
of the heat kernel, and hence in the propagators and the consequent
short-distance properties of quantum fields.   

\section{Spacetime foam?}

The idea that the ``effective infrared dimension'' might differ from four 
goes back to work by Hu and O'Connor \cite{Hu}, but the relevance to 
short-distance quantum gravity was not fully appreciated at that time.
To investigate this prospect further, though, we should better understand 
the physics underlying BKL behavior.

The BKL picture was originally developed as an attempt to understand
the cosmology of the very early Universe near an initial spacelike 
singularity.  Near such a singularity, light rays are typically very 
strongly focused by the gravitational field, leading to the collapse
of light cones and the shrinking of particle horizons.  This ``asymptotic 
silence'' \cite{HUR} is the key ingredient in the ultralocal behavior 
of the equations of motion, from which the rest of the BKL results 
follow.

Small-scale quantum gravity has no such spacelike singularity, so if 
a similar mechanism is at work, something else must account for the 
focusing of null geodesics.  An obvious candidate is ``spacetime foam,''
small-scale quantum fluctuations of geometry.  Seeing whether such
an explanation can work is very difficult; it will ultimately require 
that we understand the full quantum version of the Raychaudhuri 
equation.  As a first step, though, let us start with the classical 
Raychaudhuri equation for the focusing of null geodesics,
\begin{equation}
\frac{d\theta}{d\lambda} = -\frac{1}{2}\theta^2 
      - \sigma_\alpha{}^\beta \sigma_\beta{}^\alpha
     + \omega_{\alpha\beta}\omega^{\alpha\beta} 
      - R_{\alpha\beta}k^\alpha k^\beta .
\label{Ca14}
\end{equation}
Here, $\theta$ is the expansion of a bundle of light rays, essentially
$(1/A)({dA}/{d\lambda})$ where $A$ is the cross-sectional area; 
$\sigma$ and $\omega$ are the shear and rotation of the bundle.  
Negative terms on the right-hand side of (\ref{Ca14}) decrease
expansion, and thus focus null geodesics; positive terms contribute
to defocusing.

If we naively treat (\ref{Ca14}) as an operator equation in the 
Heisenberg picture and take the expectation value, ignoring for the
moment the need for renormalization, we see that quantum fluctuations 
in the expansion and shear should  focus geodesics.  Indeed,
\begin{equation}
\langle\theta^2\rangle = \langle\theta\rangle^2 + (\Delta\theta)^2 ,
\label{Ca15}
\end{equation}
with a similar equation for $\sigma$, so the uncertainties $\Delta\theta$
and $\Delta\sigma$ contribute negative terms to the right-hand side 
of (\ref{Ca14}).  We can estimate the size of these fluctuations by noting
that the expansion $\theta$ is roughly canonically conjugate to the 
cross-sectional area $A$ \cite{Epp}.  Indeed, $\theta$ is the trace of an 
extrinsic curvature, which is, as usual, conjugate to the corresponding 
volume element.  Keeping track of factors of $\hbar$ and $G$, we find 
an uncertainty relation of the form
\begin{equation}
\Delta{\bar\theta}\Delta A \sim \ell_p ,
\label{Ca16}
\end{equation}
where $\bar\theta$ is the expansion averaged over a Planck distance
along the congruence.  In many approaches to quantum gravity---for
instance, loop quantum gravity---we expect areas to be quantized
in Planck units.  It is thus plausible that $\Delta A\sim\ell_p^2$,
which would imply fluctuations of $\theta$ of order $1/\ell_p$.  This
would mean very strong focusing at the Planck scale, as desired.

As I have presented it, this argument is certainly inadequate.  To begin 
with, I have not specified which congruence of null geodesics I am 
considering.  In the BKL analysis, a spacelike singularity determines a 
special null congruence.  In short-distance quantum gravity, no such 
structure exists, and we will have to work hard to define $\theta$ as a 
genuine observable.  

Moreover, while the classical shear contributes a 
negative term to the right-hand side of (\ref{Ca14}), the operator 
product $\sigma_\alpha{}^\beta \sigma_\beta{}^\alpha$ in a 
quantum theory must be renormalized, and need not remain positive.  
Indeed, it is known that this quantity becomes negative near the horizon 
of a black hole \cite{Candelas}; this is a necessary consequence of
the fact that Hawking radiation decreases the horizon area.   On the 
other hand, there are circumstances in which a particular average of 
this operator over a special null geodesic must be positive to avoid 
violations of the generalized second law of thermodynamics \cite{Wall}.  
The question of whether quantum fluctuations and ``spacetime foam'' 
at the Planck scale can lead to something akin to asymptotic silence 
thus remains open.

\section{What next?}

In short, the proposal I am making is this: that spacetime foam strongly 
focuses geodesics at the Planck scale, leading to the BKL behavior 
predicted by the strongly coupled Wheeler-DeWitt equation.  If this
suggestion proves to be correct, it leads to  a novel and interesting
picture of the small-scale structure of spacetime.  At each point, the 
dynamics picks out a ``preferred'' spatial direction, leading to 
approximately (1+1)-dimensional local physics.  The preferred directions 
are presumably determined classically by initial conditions, but because 
of the chaotic behavior of BKL bounces, they are quickly randomized;
in the quantum theory, they are picked out by an initial wave function,
but again one expects evolution to scramble any initial choices.  From 
point to point, these preferred directions vary continuously, but they
oscillate rapidly \cite{Montani}.  Space at a fixed time is thus threaded 
by rapidly fluctuating lines, and spacetime by two-surfaces; the 
leading behavior of the physics is described by an approximate 
dimensional reduction to these surfaces.

There is a danger here, of course: the process I have described breaks
Lorentz invariance at the Planck scale, and even small violations at
that scale can be magnified and lead to observable effects at large
scales \cite{Mattingly}.  Note, though, that the symmetry violations 
in this scenario vary rapidly and essentially stochastically in both 
space and time.  Such ``nonsystematic'' Lorentz violations are 
harder to study, but there is evidence that they lead to much weaker 
observational constraints \cite{Basu}.

The scenario I have presented is still very speculative, but I believe 
it deserves further investigation.  One avenue might be to use results
from the eikonal approximation \cite{tHooft,Verlinde,Kabat}.  In
this approximation, developed to study very high energy scattering,
 a similar dimensional reduction takes place, with drastically disparate 
time scales in two pairs of dimensions.  Although the context is very
different, the technology developed for this approximation could 
prove useful for the study of Planck scale gravity.

\vspace{1.5ex}
\begin{flushleft}
\large\bf Acknowledgments
\end{flushleft}
 
I would like to thank Beverly Berger, David Garfinkle, and Bei-Lok Hu
for valuable advice on this project, and Leonardo Modesto and Lenny 
Susskind, among others, for asking hard questions.
This work was supported in part by U.S.\ Department of Energy grant
DE-FG02-91ER40674.


\begin{thebibliography}{99}
\bibitem{HawkingEllis} 
   S.~W.\ Hawking and G.~F.~R.\ Ellis, 
   \emph{The large scale structure of space-time}, Cambridge
   University Press, Cambridge, 1973.
\bibitem{Mattingly}
   D.~Mattingly, \emph{Living Rev.\ Relativity} \textbf{8},  5 (2005),
   eprint gr-qc/0502097.
\bibitem{Regge} 
    T.~Regge, \emph{Nuovo Cimento A} \textbf{19}, 558 (1961).
\bibitem{Williams} 
    M.~Rocek and R.~M.~Williams, \emph{Phys.\ Lett.} \textbf{B104},
    31 (1981).
\bibitem{Loll} 
    R.~Loll, \emph{Living Rev.\ Relativity} \textbf{1}, 13 (1998),
   eprint gr-qc/9805049.
\bibitem{AJL}
   J.~Ambj{\o}rn, J.~Jurkiewicz, and R.~Loll, \emph{Phys.\ Rev.\ Lett.} 
   \textbf{85}, 924 (2000), eprint hep-th/0002050.
\bibitem{AJL2}
   J.~Ambj{\o}rn, J.~Jurkiewicz, and R.~Loll, \emph{Phys.\ Rev.\ Lett.} 
   \textbf{93}, 131301 (2004), eprint hep-th/0404156.
\bibitem{AJL3} 
   J.~Ambj{\o}rn, J.~Jurkiewicz, and R.~Loll, \emph{Phys.\ Rev.}
   \textbf{D72} 064014 (2005), eprint hep-th/0505154.
\bibitem{AJL4} 
    J.~Ambj{\o}rn, J.~Jurkiewicz, and R.~Loll, \emph{Phys.\ Lett.}
    \textbf{B607}, 205 (2005), eprint hep-th/0411152.
\bibitem{Kommu} 
    R.~Kommu, in preparation.
\bibitem{spec} 
   J.~Ambj{\o}rn, J.~Jurkiewicz, and R.~Loll, \emph{Phys.\ Rev.\ Lett.} 
  \textbf{95}, 171301 (2005), eprint hep-th/0505113.
\bibitem{Synge} 
   J.~L.~Synge, \emph{Relativity: The General Theory}, North-Holland, 
   Amsterdam, 1960.
\bibitem{Henson} 
   D.~Benedetti and J.~Henson, ``Spectral geometry as
   a probe of quantum spacetime,'' eprint arXiv:0911.0401v1 [hep-th].
\bibitem{Weinberg} 
   S.~Weinberg, ``Ultraviolet divergences in quantum theories of 
   gravitation,'' in \emph{General Relativity: An Einstein Centenary 
   Survey}, edited by S.~W.~Hawking and W.~Israel, Cambridge University 
   Press, Cambridge, 1979, pp.~790--831.
\bibitem{Reuter}
   M.~Reuter and F.~Saueressig, \emph{Phys.\ Rev.} \textbf{D65}, 
   065016 (2002), eprint hep-th/0110054.
\bibitem{Litim} 
   D.~F.~Litim, \emph{Phys.\ Rev.\ Lett.} \textbf{92}, 201301 (2004),
   eprint hep-th/0312114.
\bibitem{Nieder} 
   M.~Niedermaier, \emph{Class.\ Quant.\ Grav.} \textbf{24}, R171 (2007),
   eprint gr-qc/0610018.
\bibitem{Reuter2} 
   O.~Lauscher and M.~Reuter, \emph{JHEP} \textbf{0510} 050 (2005),
   eprint hep-th/0508202.
\bibitem{Litimb} 
   D.~F.~Litim, \emph{AIP Conf.\ Proc.} \textbf{841}
   322 (2006), eprint hep-th/0606044.
\bibitem{Modesto}
   L.~Modesto, ``Fractal Structure of Loop Quantum Gravity,'' eprint
   arXiv:0812.2214 [gr-qc].
\bibitem{Atick} 
   J.~J.~Atick and E.~Witten, \emph{Nucl.\ Phys.} \textbf{B310}, 291 
   (1988).
\bibitem{Horava}
   P.~Ho{\v r}ava, \emph{Phys.\ Rev.} \textbf{D79}, 084008 (2009),
   eprint arXiv:0901.3775 [hep-th].
\bibitem{Blas} 
   D.~Blas, O.~Pujolas, and S.~Sibiryakov, \emph{JHEP}
   \textbf{0910}, 029 (2009), eprint arXiv:0906.3046 [hep-th].
\bibitem{Horava2}
   P.~Ho{\v r}ava, \emph{Phys.\ Rev.\ Lett.} \textbf{102}, 161301 (2009),
   eprint arXiv:0902.3657 [hep-th].
\bibitem{Rideout} 
   D.~Rideout,  ``Dynamics of causal sets,'' eprint arXiv:gr-qc/0212064.
\bibitem{Sorkin} 
   R.~Sorkin, personal communication.
\bibitem{Crane} 
   L.~Crane and L.~Smolin, \emph{Nucl.\ Phys.} \textbf{B267}, 714 (1986).
\bibitem{Connes} 
   A.~Connes, ``Noncommutative geometry year 2000,'' in
   \emph{Highlights of Mathematical Physics}, edited by A.~Fokas et al., 
   American Mathematical Society, 2002, eprint math.QA/0011193.
\bibitem{DeWitt}
   B.~S.~DeWitt, \emph{Phys.\ Rev.} \textbf{160}, 1113 (1967).
\bibitem{Carlip}
   S.~Carlip, \emph{Rept.\ Prog.\ Phys.} \textbf{64}, 885 (2001), eprint
   gr-qc/0108040.
\bibitem{Isham} 
   C.~J.~Isham, \emph{Proc.\ R.\ Soc.\ London} \textbf{A351}, 209 (1976).
\bibitem{Pilati} 
   M.~Pilati, ``Strong Coupling Quantum Gravity: An Introduction,'' in 
   \emph{Quantum structure of space and time}, edited by M.~J.~Duff and 
   C.~J.~Isham, Cambridge University Press, Cambridge, 1982, pp.~53--69.
\bibitem{Maeda} 
   K.~Maeda and M.~Sakamoto, \emph{Phys.\ Rev.} \textbf{D54}, 1500 
   (1996), eprint hep-th/9604150.
\bibitem{HPT} 
   M.~Henneaux, M.~Pilati, and C.~Teitelboim, \emph{Phys.\ Lett.} 
   \textbf{110B}, 123 (1982).
\bibitem{Teitelboim} 
   C.~Teitelboim, \emph{Phys.\ Rev.} \textbf{D25}, 3159 (1982).
\bibitem{Pilati2} 
   M.~Pilati, \emph{Phys.\ Rev.} \textbf{D26}, 2645 (1982); 
   \emph{Phys.\ Rev.} \textbf{D28}, 729 (1983).
\bibitem{Rovelli}
   C.~Rovelli, \emph{Phys.\ Rev.} \textbf{D35}, 2987 (1987).
\bibitem{Husain} 
   V.~Husain, \emph{Class.\ Quant.\ Grav.} \textbf{5}, 575 (1988).
\bibitem{Francisco} 
   G.~Francisco and M.~Pilati, \emph{Phys.\ Rev.} \textbf{D31}, 241 
   (1985).
\bibitem{Salopek} 
   D.~S.~Salopek, \emph{Class.\ Quant.\ Grav.} \textbf{15}, 1185 (1998),
   eprint gr-qc/9802025.
\bibitem{Kirillov} 
   A.~A.~Kirillov, \emph{Int.\ J.\ Mod.\ Phys} \textbf{D3}, 431 (1994).
\bibitem{BKL}
   V.~A.~Belinskii, I.~M.~Khalatnikov, and E.~M.~Lifshitz, 
   \emph{Adv.\ Phys.} \textbf{19}, 525 (1970); \emph{Adv.\ Phys.}
   \textbf{31}, 639 (1982).
\bibitem{HUR} 
   J.~M.~Heinzle, C.~Uggla, and N.~R{\"o}hr, \emph{Adv.\ Theor.\ Math.\ 
   Phys.} \textbf{13}, 293 (2009), eprint gr-qc/0702141.
\bibitem{Henneaux} 
   M.~Henneaux, \emph{Bull.\ Math.\ Soc.\ Belg.} 
   \textbf{31}, 47 (1979).
\bibitem{Helfer} 
   A.~Helfer et al., \emph{Gen.\ Rel.\ Grav.} \textbf{20}, 875 (1988).
\bibitem{Misner}
   C.~W.~Misner, \emph{Phys.\ Rev.\ Lett.} \textbf{22}, 1071 (1969).
\bibitem{Chernoff} 
   D.~F.~Chernoff and J.~D.~Barrow, \emph{Phys.\ Rev.\ Lett.} \textbf{50}, 
   134 (1983).
\bibitem{Kirillov2} 
   A.~A.~Kirillov and G.~Montani, \emph{Phys.\ Rev.} \textbf{D56}, 
   6225 (1997).
\bibitem{Futamase} 
   T.~Futamase,  \emph{Phys.\ Rev.} \textbf{D29}, 2783 (1984).
\bibitem{Berkin} 
   A.~L.~Berkin, \emph{Phys.\ Rev.} \textbf{D46}, 1551 (1992).
\bibitem{DeWitt2} 
   B.~S.~DeWitt, ``Dynamical theory of groups and fields,'' in 
   \emph{Relativity, Groups and Topology}, edited by C.~DeWitt and
   B.~DeWitt, Gordon and Breach, New York, 1964, pp.~587--820.
\bibitem{Gibbons} 
   G.~W.~Gibbons, ``Quantum field theory in curved spacetime,'' in 
   \emph{General Relativity: An Einstein Centenary Survey}, edited by 
   S.~W.~Hawking and W.~Israel, Cambridge University Press, Cambridge, 
   1979, pp.~639--79.
\bibitem{Vassilevich}
   D.~V.~Vassilevich, \emph{Phys.\ Rept.} \textbf{388}, 279 (2003),
   eprint hep-th/0306138.
\bibitem{Hu} 
   B.~L.~Hu and D.~J.~O'Connor,  \emph{Phys.\ Rev.} \textbf{D34}, 2535 
   (1986).
\bibitem{Epp} 
   R.~Epp, ``The Symplectic Structure of General Relativity in the 
   Double Null (2+2) Formalism,'' eprint gr-qc/9511060.
\bibitem{Candelas} 
   P.~Candelas and D.~W.~Sciama, \emph{Phys.\ Rev.\ Lett.} \textbf{38},
  1372 (1977).
\bibitem{Wall} 
   A.~Wall, ``Proving the achronal averaged null energy condition from the
   generalized second law,'' eprint arXiv:0910.5751.
\bibitem{Montani}
   G.~Montani, \emph{Class.\ Quant.\ Grav.} \textbf{12}, 2505 (1990).
\bibitem{Basu}
   S.~Basu and D.~Mattingly, \emph{Class.\ Quant.\ Grav.} \textbf{22},
   3029 (2005), eprint astro-ph/0501425.
\bibitem{tHooft}
   G.~'t Hooft, \emph{Phys. \ Lett.} \textbf{ 198B}, 61 (1987).
\bibitem{Verlinde}
   H.~L.~Verlinde and E.~P.~Verlinde,  \emph{Nucl.\ Phys.} \textbf{B371},
   246 (1992), eprint hep-th/9110017.
\bibitem{Kabat}
   D.~Kabat and M.~Ortiz, \emph{Nucl.\ Phys.} \textbf{B388}, 570 (1992),
   eprint hep-th/9203082.
\end{thebibliography}
\end{document}